# The ground-state energy and external potential as functionals of the electron density and their derivatives

Tamás Gál

Quantum Theory Project, Department of Physics, University of Florida,
Gainesville, Florida 32611, USA

**Abstract:** It is shown that the ground-state energy as a functional solely of the electron density is determined by the asymptotic value of the derivative of the degree-one homogeneous extension of the universal density functional $F[n]$ at the given electron number. This has the consequence that its derivative cannot be properly determined. Carrying out the derivative of $E[n[N,v]]$ with respect to $v(r)$ leads to a paradox, which is resolved by the non-differentiability of $E[n]$ if one follows traditional wisdom regarding the non-invertibility of the linear response function. However, considering the derivative of $v[n[v]]$ through the one-electron case shows that this paradox has a more elementary origin, namely, an unaccounted restriction of the $v(r)$ domain.

The great success of density functional theory (DFT) [1] is due to its use of the electron density $n(\vec{r})$ in the place of the many-electron wavefunction. In DFT, the ground-state energy is given by a density functional

$$E_v[n] = F[n] + \int n(\vec{r}) v(\vec{r}) d\vec{r} , \qquad (1)$$

which attains its minimum subject to the constraint

$$\int n(\vec{r}) d\vec{r} = N \qquad (2)$$

at the ground-state density corresponding to the given external potential $v(\vec{r})$. This variational principle has been established by Hohenberg and Kohn [2], and yields the Euler-Lagrange equation

$$\frac{\delta F[n]}{\delta n(\vec{r})} + v(\vec{r}) = \mu \qquad (3)$$

for the determination of the ground-state $n(\vec{r})$ in a given $v(\vec{r})$, where $\mu$ is the Lagrange multiplier corresponding to Eq.(2). The DFT method for determining the ground-state density and energy is analogous to the basic quantum mechanical scheme, where the minimum of the wavefunction functional

$$E_v[\psi] = \langle \psi | \hat{T} + \hat{V}_{ee} | \psi \rangle + \int n(\vec{r}) v(\vec{r}) d\vec{r} , \qquad (4)$$

with

$$n(\vec{r}) = \langle \psi | \hat{n}(\vec{r}) | \psi \rangle , \qquad (5)$$

subject to the constraint

$$\langle \psi | \psi \rangle = 1 \qquad (6)$$

delivers the ground-state energy, giving the Schrödinger equation as the corresponding Euler-Lagrange equation,

$$\left( \hat{T} + \hat{V}_{ee} + \sum_{i=1}^{N} v(\vec{r}_i) \right) \psi_N = E \psi_N , \qquad (7)$$

where $\psi_N$ stands for $N$-particle antisymmetric wavefunctions $\psi(\vec{r}_1,...,\vec{r}_N)$ (not denoting spin throughout for simplicity in presentation).

As the first Hohenberg-Kohn theorem [2] proves, the external potential emerges as a functional of the density, i.e. $v(\vec{r})[n]$. If for a given $n(\vec{r})$, there exists a potential $v(\vec{r})$ that yields $n(\vec{r})$ as the corresponding ground-state density (or as a linear combination of the corresponding ground-state densities [3]), the functional $F[n]$ is differentiable [4], and its derivative delivers the functional $v(\vec{r})[n]$ via Eq.(3). As long as $F[n]$ is defined only for $n(\vec{r})$'s of integer $N$, its derivative, necessarily, is an ambiguous restricted derivative $\left.\frac{\delta F[n]}{\delta n(\vec{r})}\right|_N$, determined only up to an arbitrary additive constant [5,6], implying that $v(\vec{r})[n]$ is determined by Eq.(3) also only up to an additive constant. If the ground-state energy $E[N,v]$ is generalized for fractional electron number $N$ (e.g., by the zero-temperature grand canonical ensemble scheme [7]), the unrestricted derivative of $F[n]$ can be used in Eq.(3), which determines the Lagrange multiplier $\mu$ uniquely as

$$\mu = \left(\frac{\partial E[N,v]}{\partial N}\right)_v, \tag{8}$$

giving $\mu$ the interpretation as an electronic chemical potential [8]. Then, $v(\vec{r})[n]$ emerges as

$$v(\vec{r})[n] = -\frac{\delta F[n]}{\delta n(\vec{r})} + \mu[n]. \tag{9}$$

$\mu[n]$ cannot be obtained through Eq.(8), by simply substituting

$$N = N[n] \equiv \int n(\vec{r})\,d\vec{r} \tag{10}$$

and $v = v[n]$ [9]. This would only give

$$\mu[n] = \frac{\partial E\left[N, -\frac{\delta F[n]}{\delta n(\vec{r})}\right]}{\partial N} + \mu[n]; \tag{11}$$

that is,

$$\frac{\partial E\left[N, -\frac{\delta F[n]}{\delta n(\vec{r})}\right]}{\partial N} = 0 \tag{12}$$

(which, though, is an important result in itself, saying that by choosing $v(\vec{r})$ precisely as $F[n]$'s derivative, the energy becomes constant in $N$). $\mu[n]$ is determined by the fixation of the constant ambiguity of the external potential [9], which, in the case of Coulombic systems, is $v(\infty) = 0$. This then yields

$$\mu[n] = \frac{\delta F[n]}{\delta n(\infty)}. \tag{13}$$

In the DFT literature, it is a basic statement that due to the first Hohenberg-Kohn theorem, the ground-state density determines all properties of an electronic system, in particular the ground-state energy, as both inputs of the Schrödinger equation, $N$ and $v(\vec{r})$, are determined by the density. It is, therefore, worth examining the nature of the external potential and the ground-state energy as sole density functionals.

The ground-state energy $E[N,v]$ may be obtained formally as a sole density functional by substituting Eqs.(9) and (10) for $N$ and $v(\vec{r})$. This, however, does not tell us too much, hiding the essence of this $n(\vec{r})$-dependency. Therefore, we write the functional $F[n]$ in the form

$$F[n] = F_{\int n}[n] \,, \tag{14}$$

where $F_N[n]$ is such that

$$F_N[\alpha n] = \alpha F_N[n] \,, \tag{15}$$

i.e. $F_N[n]$ is homogeneous of degree one in $n(\vec{r})$. Writing $F[n]$ in such a form can be done without harming generality; further, $F_N[n]$ is determined uniquely by the degree-one homogeneity requirement [10] (see Appendix). The inverse connection between $F_N[n]$ and $F[n]$ is given by [10]

$$F_N[n] = \left(\frac{\int n}{N}\right) F\left[N \frac{n}{\int n}\right] \,. \tag{16}$$

Since $F_N[n]$ (with $N$ fixed) and $F[n]$ equals for $n(\vec{r})$'s of a given $N$ (which is given in $F_N[n]$'s index), their derivatives, for $n(\vec{r})$'s of $N$, differ only by an ($\vec{r}$-independent) constant [6]. Concretely,

$$\frac{\delta F_N[n]}{\delta n(\vec{r})} = \frac{\delta F[n]}{\delta n(\vec{r})} - \frac{1}{\int n(\vec{r}')d\vec{r}'}\left(\int n(\vec{r}')\frac{\delta F[n]}{\delta n(\vec{r}')}d\vec{r}' - F_N[n]\right) \,. \tag{17}$$

With this $F_N[n]$, then, the Euler-Lagrange equation Eq.(3) will emerge as

$$\frac{\delta F_N[n]}{\delta n(\vec{r})} + v(\vec{r}) = \frac{E}{N} \,, \tag{18}$$

which can be checked by multiplying Eq.(18) by $n(\vec{r})$ and integrating it.

From Eq.(18), $v(\vec{r})[n]$ emerges as

$$v(\vec{r})[n] = -\frac{\delta F_N[n]}{\delta n(\vec{r})} + \frac{E[n]}{N} \,. \tag{19}$$

Again, the sole density functional $E[n]$ cannot be obtained by substituting Eq.(19) in

$$E[n] \equiv E[N[n],v[n]] \equiv E_{v[n]}[n] \equiv F_{\int n}[n] + \int n(\vec{r})v(\vec{r})[n]d\vec{r} \,, \tag{20}$$

since this would be circular logic, yielding an identity, which can be seen by utilizing the property

$$\int n(\vec{r}) \frac{\delta F_N[n]}{\delta n(\vec{r})} d\vec{r} = F_N[n] \ , \qquad (21)$$

holding for degree-one homogeneous functionals. Instead, as $E/N$ enters Eq.(18) as a Lagrange multiplier, similar to $\mu[n]$, it will be given as a density functional by the asymptotic fixation of $v(\vec{r})$. That is,

$$E[n] = N \frac{\delta F_N[n]}{\delta n(\infty)} \ , \qquad (22)$$

which means that the ground-state energy is determined as a sole functional of the density by the asymptotic value of the derivative of the degree-one homogeneous extension of $F[n]$ multiplied by $N$ (of course, the explicit $N$'s appearing in this expression should be considered as density functionals through Eq.(10)).

It is important to recognize that the above determination of $E[n]$ is completely analogous to how the ground-state energy can be obtained as a sole wavefunction functional in quantum mechanics. Consider the one-particle case for simplicity. The Schrödinger equation determines $v(\vec{r})$ as a functional of $\psi(\vec{r})$, as

$$v(\vec{r})[\psi] = -\frac{\hat{T}\psi}{\psi} + E[\psi] \ . \qquad (23)$$

Substituting Eq.(23) in

$$E[\psi] \equiv E_{v[\psi]}[\psi] \equiv \langle \psi | \hat{T} | \psi \rangle + \int n(\vec{r}) v(\vec{r})[\psi] d\vec{r} \qquad (24)$$

gives nothing but an identity. Therefore, in Eq.(23), $E[\psi]$ has to be determined by the asymptotic fixation of $v(\vec{r})$. This yields

$$E[\psi] = \frac{\hat{T}\psi(\infty)}{\psi(\infty)} \ . \qquad (25)$$

It is absolutely not surprising that the energy, either as a functional of the wavefunction or as a number itself, is determined by the fixation of $v(\vec{r})$'s ambiguity, i.e. by the choice of the zero of energy. As we have seen above, similar happens in DFT. In the case of more than one particle, $v(\vec{r})[\psi]$ arises by applying $v(\vec{r}_i \to \infty) = 0$ for all $i$ but one in the Schrödinger equation divided by $\psi$.

Having $v(\vec{r})$, together with $\mu$ and $E$, determined as a density functional, we may calculate their derivatives with respect to the density. With this, however, an essential paradox arises. Using the basic fact

$$\left(\frac{\delta E[N,v]}{\delta v(\vec{r})}\right)_N = n(\vec{r}) \tag{26}$$

from perturbation theory, we have

$$\int \left(\frac{\delta E[N,v]}{\delta v(\vec{r})}\right)_N d\vec{r} = N \ . \tag{27}$$

Now, applying the chain rule of differentiation for $E[N,v] \equiv E[n[N,v]]$, we obtain

$$\iint \frac{\delta E[n]}{\delta n(\vec{r}\,')} \left(\frac{\delta n(\vec{r}\,')}{\delta v(\vec{r})}\right)_N d\vec{r}\,' d\vec{r} = N \ , \tag{28a}$$

or, with the use of Eq.(26) again,

$$\iint \frac{\delta E[n]}{\delta n(\vec{r}\,')} \left(\frac{\delta E[N,v]}{\delta v(\vec{r}\,')\delta v(\vec{r})}\right)_N d\vec{r}\,' d\vec{r} = N \ . \tag{28b}$$

The paradox becomes apparent when one utilizes the symmetry of second derivatives in the indices of variables, which leads to

$$\int \frac{\delta E[n]}{\delta n(\vec{r}\,')} \int \left(\frac{\delta n(\vec{r})}{\delta v(\vec{r}\,')}\right)_N d\vec{r} d\vec{r}\,' = N \ , \tag{29a}$$

i.e.

$$\int \frac{\delta E[n]}{\delta n(\vec{r}\,')} \left(\frac{\delta N}{\delta v(\vec{r}\,')}\right)_N d\vec{r}\,' = N \ , \tag{29b}$$

which, however, should be zero. A similar paradox can be arrived at for $\mu[n]$, too. Utilizing the Maxwell relation

$$\left(\frac{\delta \mu[N,v]}{\delta v(\vec{r})}\right)_N = \left(\frac{\partial n(\vec{r})}{\partial N}\right)_v \tag{30}$$

and the simple fact that the Fukui function $\left(\frac{\partial n(\vec{r})}{\partial N}\right)_v$ [11], an essential reactivity indicator of conceptual DFT [12], integrates to one, we have

$$\int \left(\frac{\delta \mu[N,v]}{\delta v(\vec{r})}\right)_N d\vec{r} = 1 \ . \tag{31}$$

Now, applying the chain rule for $\mu[N,v] \equiv \mu[n[N,v]]$, we obtain

$$\iint \frac{\delta\mu[n]}{\delta n(\vec{r}')}\left(\frac{\delta n(\vec{r}')}{\delta v(\vec{r})}\right)_N d\vec{r}'d\vec{r} = 1 \ . \tag{32}$$

The paradox, again, is apparent when one utilizes the symmetry of $\left(\frac{\delta n(\vec{r}')}{\delta v(\vec{r})}\right)_N$ in $\vec{r}$ and $\vec{r}'$; consequently,

$$\int\left(\frac{\delta E[N,v]}{\delta v(\vec{r}')\delta v(\vec{r})}\right)_N d\vec{r} = 0 \ . \tag{33}$$

This paradox disappears if (i) $E[n]$ and $\mu[n]$ cannot be differentiated with respect to $n(\vec{r})$, or (ii) $\left(\frac{\delta E[N,v]}{\delta v(\vec{r}')\delta v(\vec{r})}\right)_N$ is not symmetric in its variables, contrary to widely held view. (The latter would make the causality paradox of time-dependent DFT [13,14] disappear, too.) It seems though that the first alternative resolves the paradox, since as pointed out in [9], the restriction of the density domain to densities corresponding to potentials with $v(\infty)=0$ adds a non-negligible correction to the derivative of $\mu[n]$ that makes its determination infeasible. The argument given in [9] for $\mu[n]$ manifests itself in a sharper form in the case of $E[n]$, since one may rely on the one-electron case directly on the basis of Eq.(18), with $F_1[n] \equiv T_W[n]$, the Weizsäcker functional

$$T_W[n] = \frac{1}{8}\int\frac{|\nabla n(\vec{r})|^2}{n(\vec{r})}d\vec{r} \ , \tag{34}$$

which is *the* exact degree-one homogeneous one-particle *F* functional [10]. Eq.(18) thus gives

$$\frac{\delta T_W[n]}{\delta n(\vec{r})} + v(\vec{r}) = E \tag{35}$$

for *N*=1, where

$$\frac{\delta T_W[n]}{\delta n(\vec{r})} = \frac{1}{8}\left(\frac{\nabla n(\vec{r})}{n(\vec{r})}\right)^2 - \frac{1}{4}\frac{\nabla^2 n(\vec{r})}{n(\vec{r})} \ . \tag{36}$$

For one-particle densities $n_1(\vec{r})$, then, we have

$$E[n_1] = \frac{\delta T_W[n_1]}{\delta n(\infty)} \ . \tag{37}$$

The problem arises when one differentiates this expression with respect to $n(\vec{r})$ and finds it disappearing due to the entering delta functions $\delta(\vec{r}-\infty)$ – while the ground-state energy obviously does depend on the density. Note that the fact that we use an expression that is valid only for one-particle densities brings only a constant ambiguity to the derivative,

$$\frac{\delta E[n_1]}{\delta n(\vec{r})} = \frac{\delta^2 T_W[n_1]}{\delta n(\vec{r})\delta n(\infty)} + c[n_1] \ , \tag{38}$$

which cannot account for this contradiction. The resolution emerges from the fact that Eq.(22) (and Eq.(13)) is valid only for densities that correspond to $v(\vec{r})$'s with $v(\infty) = 0$. The restriction $v[n](\infty) = 0$ on the density domain may not seem to be a severe one; however, (i) it brings an ambiguity of $+\lambda \frac{\delta v[n](\infty)}{\delta n(\vec{r})}$ to derivatives with respect to $n(\vec{r})$, and (ii) in addition, it excludes potentials going to infinity as $\vec{r} \to \infty$. Thus, we only have

$$\frac{\delta E[n]}{\delta n(\vec{r})} = N\frac{\delta^2 F_N[n]}{\delta n(\vec{r})\delta n(\infty)} + N\frac{\delta v[n](\infty)}{\delta n(\vec{r})} + const. \ . \tag{39}$$

Is this indeed the solution to the paradox of Eq.(29)? To see more clearly, exhibit the paradox more sharply by starting from

$$\int \frac{\delta v(\vec{r}^*)}{\delta v(\vec{r})} d\vec{r} = 1 \ . \tag{40}$$

Applying the chain rule for $v \equiv v[n[v]]$, we obtain

$$\iint \frac{\delta v(\vec{r}^*)[n]}{\delta n(\vec{r}')} \left(\frac{\delta n(\vec{r}')}{\delta v(\vec{r})}\right)_N d\vec{r}'d\vec{r} = 1 \ , \tag{41}$$

giving a contradiction when one utilizes Eq.(33). Utilizing Eq.(19) formally in Eq.(41) and considering Eq.(29), everything seems to be consistent, the paradox stemming from the derivative of $E[n]$ solely. For this, one only has to accept that the integral of the derivative of $\frac{\delta F_N}{\delta n(\vec{r})}[n[N,v]]$ with respect to $v(\vec{r})$ vanishes. This may seem to be a too bold expectation; however, notice that this must be the case for any differentiable functional $D[n]$ of the density if second functional derivatives are symmetric in their arguments, since $\iint \frac{\delta D[n]}{\delta n(\vec{r}')}\left(\frac{\delta n(\vec{r}')}{\delta v(\vec{r})}\right)_N d\vec{r}'d\vec{r}$ must then vanish. But, why not… ? The result that $\frac{\delta v(\vec{r})[n]}{\delta n(\vec{r}')}$ does not exist, i.e. the linear response function $\left(\frac{\delta n(\vec{r}')}{\delta v(\vec{r})}\right)_N$ is not invertible, of course would not be a surprising result, as this is precisely what is generally concluded from Eq.(33) (see Appendix of [15], e.g., for a short summary). However, the resolution of the paradox of Eq.(41) is more elementary.

Since the external potentials in $v \equiv v[n[v]]$ are restricted as $v(\infty) = 0$, the derivative $\frac{\delta v(\vec{r}')}{\delta v(\vec{r})}$ will be ambiguous, not giving Eq.(40). To show that this ambiguity is not negligible, we resolve the constraint on the potentials by introducing

$$v(\vec{r})[v] \doteq v(\vec{r}) - \int \delta(\vec{r}' - \infty) v(\vec{r}') d\vec{r}' \ . \tag{42}$$

Then, we can differentiate $v[v]$ fully, obtaining

$$\int \frac{\delta v(\vec{r}^*)[v]}{\delta v(\vec{r})} d\vec{r} = 0 \tag{43}$$

in the place of Eq.(40). With this, Eq.(41) is resolved.

It is worth generalizing the fixation of the external potential's ambiguity to get more insight. Here, we are not interested in obtaining the physical ground-state energy as a functional of the density; so we may choose the following fixation of the external potentials' constant ambiguity:

$$\int g(\vec{r}) v(\vec{r})[n] d\vec{r} = 0 \ , \tag{44}$$

where $g(\vec{r})$ is some fixed function that integrates to one and tends fast to zero with $\vec{r} \to \infty$. Note that the fixation of $v(\vec{r})[n]$ is important to have a well-defined functional, though the energy $E[N, v]$ is naturally defined for any $v(\vec{r})$. With Eq.(44), we have

$$v(\vec{r})[v] \doteq v(\vec{r}) - \int g(\vec{r}') v(\vec{r}') d\vec{r}' \tag{45}$$

instead of Eq.(42). Further, in the place of Eq.(19) with Eq.(22), we obtain the functional

$$v(\vec{r})[n] = -\frac{\delta F_N[n]}{\delta n(\vec{r})} + \int g(\vec{r}') \frac{\delta F_N[n]}{\delta n(\vec{r}')} d\vec{r}' \ , \tag{46}$$

which is obviously differentiable if $\frac{\delta F_N[n]}{\delta n(\vec{r})}$ is differentiable. Eq.(45) gives

$$\int \frac{\delta v(\vec{r}^*)[v]}{\delta v(\vec{r})} d\vec{r} = 0 \ , \tag{47}$$

which, with an application of the chain rule, yields

$$\iint \frac{\delta v(\vec{r}^*)[n]}{\delta n(\vec{r}')} \left( \frac{\delta n(\vec{r}')}{\delta v(\vec{r})} \right)_N d\vec{r}' d\vec{r} = 0 \ . \tag{48}$$

Now, we may rely on the one-particle case to find a differentiable $v(\vec{r})[n]$. Inserting $F_1[n] = T_W[n]$ in Eq.(46), we have an explicit density functional $v(\vec{r})[n]$ for one-particle densities that is differentiable. The ($\vec{r}'$-independent) constant difference of this $v(\vec{r})[n]$'s derivative from the general $v(\vec{r})[n]$'s derivative (with respect to $n(\vec{r}\,')$) cancels in Eq.(48) due to the *N*-conservation constraint on $n(\vec{r}\,')$ in the outer derivative $\left(\dfrac{\delta n(\vec{r}\,')}{\delta v(\vec{r})}\right)_N$ (see Appendix in [6] in general); thus, it does not have an effect on the right side's value in Eq.(48). Of course, the emerging one-particle $v(\vec{r})[n]$ functional will be defined for a much wider range of densities than ground-state densities (allowing practically any one-particle $n(\vec{r})$, which will correspond to some eigenstate of the arising $v(\vec{r})$), while $n(\vec{r})[v]$ in $\left(\dfrac{\delta n(\vec{r}\,')}{\delta v(\vec{r})}\right)_N$ gives only ground-state $n(\vec{r})$'s, but this does not cause any problem, since a more general inner derivative is always allowed. We have thus presented an example where a non-existence of $\dfrac{\delta v(\vec{r})[n]}{\delta n(\vec{r}\,')}$ is not an obstacle in examining Eq.(48). Note that even if we re-define $E[N,v]$ by inserting Eq.(45) into its $v(\vec{r})$ argument (i.e. fixing an energy zero accordingly), its second derivative with respect to $v(\vec{r})$ will still be $\left(\dfrac{\delta n(\vec{r}\,')}{\delta v(\vec{r})}\right)_N$, since the extra term $-Ng(\vec{r}\,')$ appearing beside $n(\vec{r}\,')$ in the first derivative vanishes by further differentiation with respect to $v(\vec{r})$.

It has thus been shown that the non-differentiability of $v(\vec{r})[n]$ with the usual fixation of its asymptotic constant comes purely from the non-differentiability of $E[n]$; however, the cause for the paradoxes of Eqs.(28) and (32) is not this fact but simply a restriction of the $v(\vec{r})$-domain not accounted for by Eq.(40).

For completeness, we mention that the td version of the Euler-Lagrange equation Eq.(3) is

$$\frac{\delta A_{\text{int.}}[n]}{\delta n(\vec{r},t)} + v(\vec{r},t) = \mu(t) \;, \qquad (49)$$

the Lagrange multiplier corresponding to the constraint

$$\int n(\vec{r},t)\,d\vec{r} = N \;, \qquad (50)$$

and as such, emerging as

$$\mu(t) = \left(\frac{\delta A[N,v]}{\delta N(t)}\right)_v \;, \qquad (51)$$

provided the action functional is defined for noninteger electron number that may change with time. From Eq.(49), then, $v[n]$ arises as

$$v(\vec{r},t)[n] = -\frac{\delta A_{\text{int.}}[n]}{\delta n(\vec{r},t)} + \mu(t)[n] ,\qquad(52)$$

where $\mu(t)[n]$ is determined by the fixation of $v(\vec{r},t)$'s ambiguity, e.g. as

$$\mu(t)[n] = \int g(\vec{r}) \frac{\delta A_{\text{int.}}[n]}{\delta n(\vec{r},t)} d\vec{r} .\qquad(53)$$

References to be added.